\documentclass[fleqn,usenatbib,useAMS]{mnras}
\usepackage{newtxtext,newtxmath}
\usepackage[T1]{fontenc}
\usepackage{ae,aecompl}
\usepackage{graphicx,subfigure}	
\usepackage{grffile}
\usepackage{amsmath}	
\usepackage{amssymb}	

\usepackage{pslatex}

\title[Project VeSElkA: HD~53929 and HD~63975]{Project VeSElkA: results of abundance analysis for HD~53929 and HD~63975
\protect\thanks{Based on observations obtained at Canada-France-Hawaii Telescope (CFHT) which is operated by the National Research Council of Canada, the Institut National des Sciences de l'Univers of the Centre National de la Recherche Scientifique of France, and the University of Hawaii. The operations at Canada-France-Hawaii Telescope are conducted with care and respect from the summit of Maunakea which is a significant cultural and historic site.}}

\author[Ndiaye et al.]
       { M.L. Ndiaye\thanks{E-mail: emn7955@umoncton.ca}, F. LeBlanc, V. Khalack \\
          D\'epartement de Physique et d'Astronomie, Universit\'e de Moncton, Moncton, N.-B., E1A 3E9, Canada\\
            }
\date{Accepted ???.
      Received ???;
      in original form ???}

\pubyear{2017}

\begin{document}
\label{firstpage}
\pagerange{\pageref{firstpage}--\pageref{lastpage}}
\maketitle

\begin{abstract}
Project VeSElkA (Vertical Stratification of Element Abundances) has been initiated with the aim to detect and study the vertical stratification of element abundances in the atmosphere of chemically peculiar stars. Abundance stratification occurs in hydrodynamically stable stellar atmospheres due to the migration of the elements caused by atomic diffusion. Two HgMn stars, HD~53929 and HD~63975 were selected from the VeSElkA sample and analysed with the aim to detect some abundance peculiarities employing the ZEEMAN2 code.
We present the results of abundance analysis of HD~53929 and HD~63975 observed recently with the spectropolarimeter ESPaDOnS at Canada-France-Hawaii Telescope. Evidence of phosphorus vertical stratification was detected in the atmosphere of these two stars. In both cases, phosphorus abundance increases strongly towards the superficial layers.
The strong overabundance of Mn found in stellar atmosphere of both stars confirms that they are HgMn type stars.
\end{abstract}


\section{Introduction}
\label{intro}
The HgMn stars constitute a subgroup of the chemically peculiar (CP) stars of the main sequence. They are characterized spectroscopically by a very large overabundance of mercury and manganese. They constitute the third CP star group of the classification done by Preston in (\citeyear{Preston1974}). Indeed, the overabundance of mercury can vary up to 6 dex (e.g. Heacox \citeyear{Heacox1979}; Cowley et al. \citeyear{Cowley+06}) and up to 3 dex for manganese (e.g. Aller \citeyear{Aller1970}; Morel et al. \citeyear{Morel+13}), both relative to their solar value. Chemical elements such as Be, P, Cu, Ga, Sr, Y, Zr, Yb, Pt, Bi are often overabundant in the atmospheres of HgMn stars, in fact their overabundance is sometimes greater than 2 dex (e.g. Takada-Hidai \citeyear{Takada-Hidai1991}, Smith \citeyear{Smith1993}, Castelli \& Hubrig \citeyear{CH+04}). In addition to these overabundant elements, He, N, Mg, Al, Co, Ni and Zn are generally deficient in HgMn stars but their underabundance often does not exceed 0.5 dex (Takada-Hidai \citeyear{Takada-Hidai1991}).

Although several studies have attempted to detect the presence of a magnetic field (e.g. Mathys \& Hubrig \citeyear{Mathys+Hubrig95}; Hubrig et al. \citeyear{HCW+99}; Hubrig \& Castelli \citeyear{Hubrig+Castelli+01}; Hubrig et al. \citeyear{HGI+12}), these stars are still considered in the literature as non-magnetic stars or at least very weakly magnetic stars.

The HgMn stars are characterized by a relatively low rotational velocity ($V$sin$i$ < 100 km s$^{-1}$, Wolff \& Preston \citeyear{Wolff+Preston78}) and have effective temperatures between 10000 and 16000 K (e.g. Smith \citeyear{Smith1996a}). There should not be a hydrogen convective zone near the surface of HgMn stars for the simple reason that these stars have high effective temperatures (e.g. Michaud et al. \citeyear{Michaud15}). Their low rotational velocity means that the atmosphere of these stars might be hydrodynamically stable, which would allow the process of atomic diffusion (Michaud \citeyear{Michaud70}) to take place there and determine the chemical properties of the outer layers of HgMn stars. Atomic diffusion can cause a vertical abundance stratification of elements present in the stellar atmosphere of HgMn stars, which in turn can change their physical structure (e.g. Hui-Bon-Hoa et al. \citeyear{HBH+00}, LeBlanc et al. \citeyear{LeBlanc+09}).

Several studies have revealed signs of vertical stratification of metals in the stellar atmosphere of certain HgMn stars. Vertical stratification of Mn was detected by Alecian (\citeyear{Alecian+1982}) in the atmosphere of $\nu$ Her and then by Sigut (\citeyear{Sigut+2001}) in the atmospheres of HD~186122 (46~Aql) and HD~179761. Vertical stratification of Cr was discovered by Savanov \& Hubrig (\citeyear{Savanov+Hubrig03}) in a sample of 10~HgMn stars who found its abundance increase towards the upper atmospheric layers in all studied stars except for HD~49606.
Thiam et al. (\citeyear{Thiam+10}) have found a vertical stratification of Mn in HD~178065, while Catanzaro et al. (2016) found evidence of Mg, Si and P stratification in the HgMn star HD~49606.

In addition to the aforementioned chemical anomalies, HgMn stars also exhibit isotopic anomalies on their surface. For instance, in some HgMn stars (such as $\chi$ Lupi), the relative abundances of the Hg and Pt isotopes are very different to their solar values (White et al. \citeyear{White+76}; Dworetsky et al. \citeyear{Dworetsky+84}; Leckrone et al. \citeyear{Leckrone+91}, \citeyear{Leckrone+93}). These isotopic anomalies are probably due to the physical process of light-induced-drift (see Atutov \& Shalagin \citeyear{Atutov+Shalagin+88} and LeBlanc \& Michaud \citeyear{LeBlanc+Michaud+93}, for more details).

Recently, a research project named VeSElkA was initiated by Khalack \& LeBlanc (\citeyear{Khalack+LeBlanc15a}, \citeyear{Khalack+LeBlanc15b}) to search for vertical stratification of element abundances in stellar atmospheres of main sequence CP stars. More than 50 slowly rotating objects with $V$sin$i$ < 40 km s$^{-1}$ were selected from the catalogue of Ap, HgMn and Am stars of Renson \& Manfroid (\citeyear{Renson+Manfroid09}) and observed with the spectropalariometer ESPaDOnS at the Canada-France-Hawaii Telescope in the frame of this project. Project VeSElkA aims to identify vertical stratification of elements in a large number of CP stars in order to guide the theoretical modeling of their atmosphere and to search for correlations between the global stellar parameters and the detected vertical abundance stratification. 
Several stars from this sample have already been analyzed and some of them revealed signatures of vertical stratification of metals. HD~22920 shows vertical stratification of Cr (Khalack \& Poitras \citeyear{Khalack+Poitras+15}), HD~95608 and HD~116235 show vertical stratification of Fe and Cr (LeBlanc et al. \citeyear{LeBlanc+15}), while HD~41076 and HD~148330 show vertical stratification of several metals (Khalack et al. \citeyear{Khalack+17}).

The purpose of this paper is to perform a spectral analysis of two stars (HD~53929 and HD~63975) identified as HgMn-type by Renson \& Manfroid (\citeyear{Renson+Manfroid09}). These two stars are part of the VeSElkA sample, each exhibiting a slow axial rotation, thus possessing a hydrodynamically stable stellar atmosphere, where atomic diffusion can be effective. Smith \& Dworetsky (\citeyear{Smith+Dworetsky93}) estimated the abundance of iron-peak elements Cr, Mn, Fe, Co and Ni in HD~53929 using IUE spectra.  Using these same spectra, Smith (1993) obtained its abundances for Mg, Al and Si, Smith (1994) estimated its Cu and Zn abundances, Smith (1996b) acquired its Ga abundance, while Smith (\citeyear{Smith1997}) evaluated its Hg abundance. For HD~63975, the only element for which we have found an abundance estimation is Hg that was obtained by Woolf \& Lambert (\citeyear{Woolf+Lambert99}). In this paper, recently obtained spectra of HD~53929 and HD~63975 are studied to perform an abundance analysis of their stellar atmospheres for a large number of elements.
\begin{table}
	\caption{Observation data of HD 53929 and HD 63975.}
	\label{tab:parameters}
	\begin{tabular}{ l l l c c }
		\hline
		Star   &  Date &  HJD & t$_{exp}$ & S/N \\
		       & (UTC) & (2400000+) & (s) & Stokes I/V
		    \\\hline
		HD 53929 & Jan 1, 2013 & 56293.95962 & 1160 &  1070/870              \\
		HD 63975 & Feb 22, 2016  & 57440.81245 & 348 &  520/420              \\\hline
	\end{tabular}
\end{table}

In Section~\ref{obs}, we provide a brief description of the acquired spectra and the reduction procedure.
The method used for the abundance analysis and the obtained results are presented in Section~\ref{spectral}, followed by the discussion in Section~\ref{discus}.

\section{Observations and data reduction}
\label{obs}
High-resolution ($R$ = $\lambda$ / ($\Delta \lambda$) = 65000) Stokes IV spectra with a high signal-to-noise ratio were obtained using ESPaDOnS (Echelle SpectroPolarimetric Device for the Observation of Stars) at the Canada-France-Hawaii Telescope. ESPaDOnS is a bench-mounted high-resolution echelle spectropolarimeter designed to obtain a complete optical spectrum (i.e. from 3700 to 10500 \text{\AA}) in a single exposure. It is able to measure all  polarization components of the stellar light (both circular and linear) with the same resolving power (Donati \citeyear{Donati+03}). The spectra were reduced with the help of the data reduction package Libre-ESpRIT developed by Donati et al. (\citeyear{Donati+97}). In the case of HgMn stars, the Stokes V spectrum shows no detectable signal because they are most probably non-magnetic stars. Thus, within this framework, we only use the Stokes I spectra for our analysis. Table~\ref{tab:parameters} presents the journal of spectral observations, where the first column gives the star's name, the second and the third columns provide respectively the date and heliocentric Julian date of observation, the fourth and fifth columns present the exposure time and the signal-to-noise ratio for the Stokes I and V spectra.

\section{Spectral analysis and results}
\label{spectral}

\subsection{Analysis method}
\label{method}
The abundance analysis of stellar atmospheres of HD~53929 and HD~63975 has been carried out in several steps. The first step was to identify all absorption lines available in the observed spectra using the NIST (Kramida et al. \citeyear{Kramida+15}; NIST ASD Team \citeyear{Kramida+15}) and VALD3 (Piskunov et al. \citeyear{Piskunov+95}; Ryabchikova et al. \citeyear{Ryab+97}; Kupka et al. \citeyear{Kupka+99}; Kupka et al. \citeyear{Kupka+00}; Ryabchikova et al. \citeyear{Ryab+15}) databases. The damping constants are taken from the VALD3 database, and the $\log$ (\textit{gf}) values of this database are used in the majority of our simulations. When unavailable in the VALD3 database, the $\log$ (\textit{gf}) values provided by NIST were employed.

\begin{figure*}
	\begin{tabular}{cc}
		\includegraphics[scale=0.65,angle=-90.]{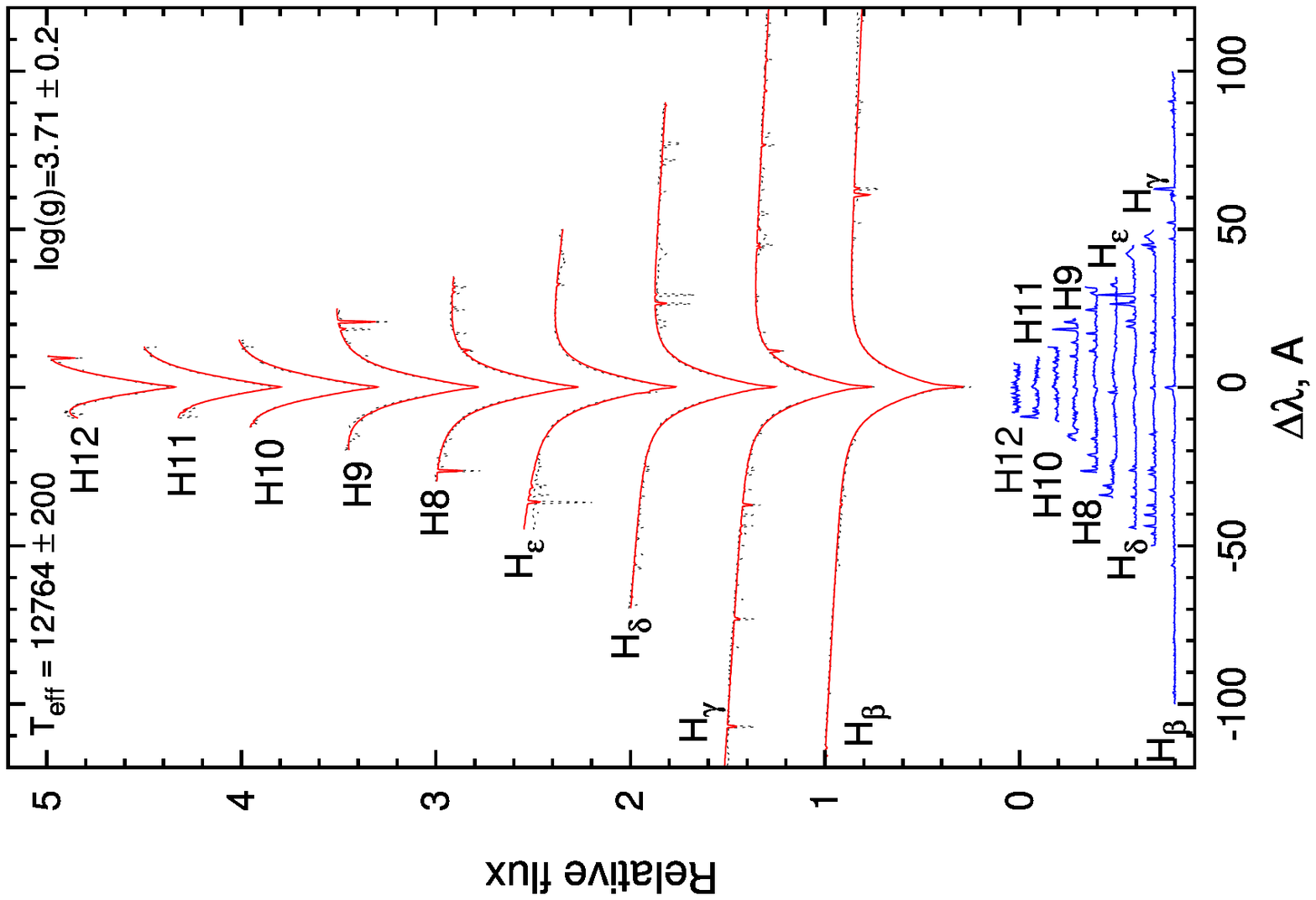} &
		\includegraphics[scale=0.65,angle=-90.]{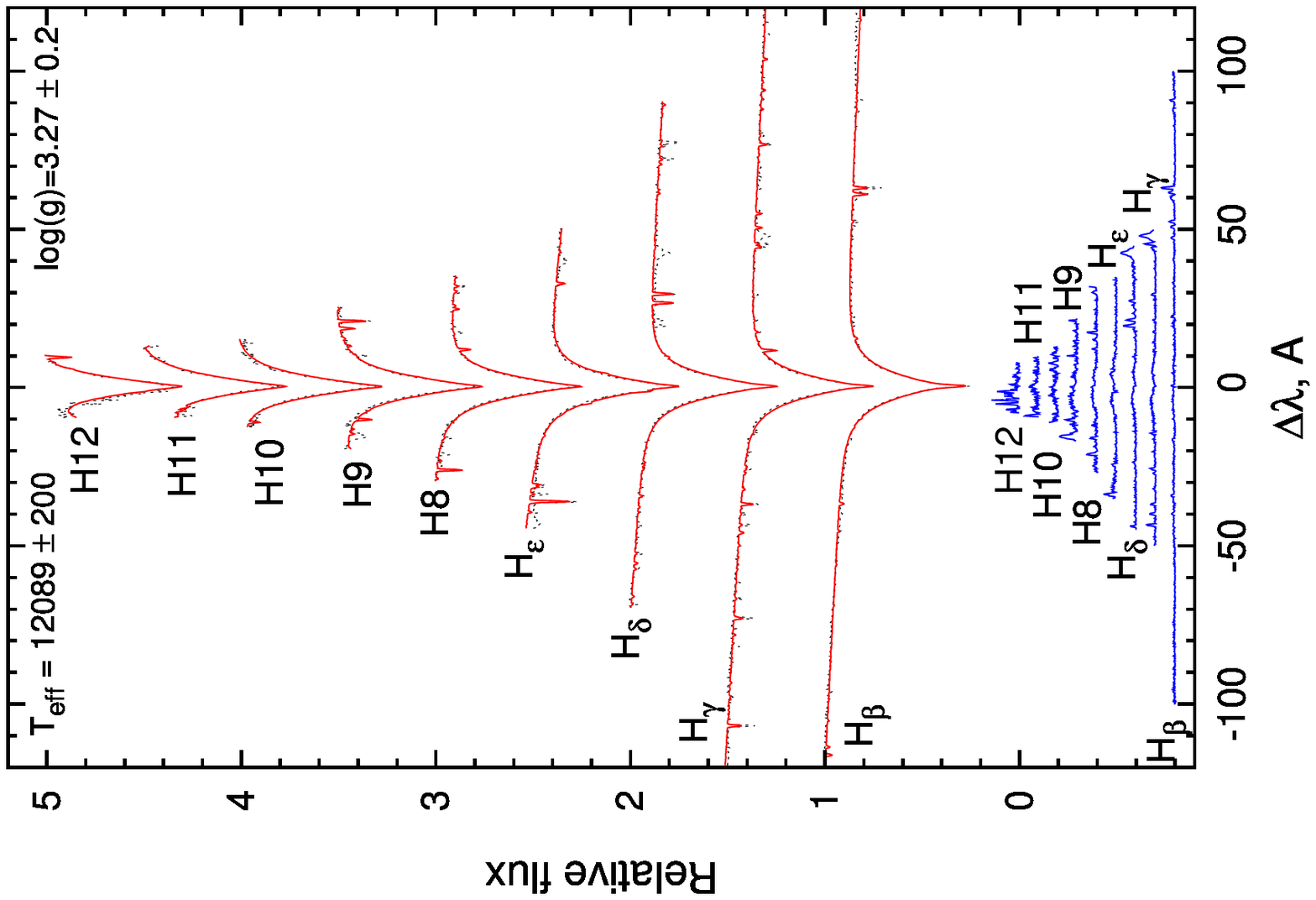}  \\
	\end{tabular}
	\caption{The Balmer line profiles in the observed spectra (thick line) of HD~53929 (right) and HD~63975 (left) are relatively well adjusted by the synthetic spectra (thin dotted line) resulting in $T_{\rm eff}$= 12764 K, $\log$(\textit{g})= 3.71 for HD~53929 and $T_{\rm eff}$= 12089 K, $\log$(\textit{g})= 3.27 for HD~63975. To visualise the quality of fits the differences between the observed and synthetic spectra are shown at the bottom of each image. For the sake of visibility the Balmer line profiles are shifted by 0.5 and the differences are shifted by 0.1.}
	\label{fig:balmer}
\end{figure*}

For each star, the physical parameters used for our simulations are presented in Table~\ref{tab:velocities}. These parameters were estimated for HD~53929 and HD~63975 with the help of FITSB2 code (Napiwotzki \citeyear{Napiwotzki+04}) by fitting the Balmer line profiles in the non-normalised spectra of studied stars with non-normalised theoretical fluxes (Khalack~\&~LeBlanc \citeyear{Khalack+LeBlanc15b}). For both stars the best fits are shown in Fig.~\ref{fig:balmer}. For each Balmer line the fit quality is presented at the bottom of each image in the form of differences between the observed and synthetic spectra. 
Stellar atmosphere models with the corresponding fluxes were computed for the studied stars using version 15 of PHOENIX code (Hauschildt et al. \citeyear{Hauschildt+97}) and the values of $T_{\rm eff}$, $\log$(\textit{g}) and metallicity listed in Table~\ref{tab:velocities}. The uncertainties given for the basic parameters of the stars shown in Table~\ref{tab:velocities} were estimated similarly to Khalack \& LeBlanc (\citeyear{Khalack+LeBlanc15b}).

\begin{table*}
\begin{center}
	\caption{Physical parameters found for HD 53929 and HD 63975.}
	\label{tab:velocities}
	\begin{tabular}{lllccccc}
\hline
Star    &   \multicolumn{2}{c}{$T_{\rm eff}$ (K)}         &$\log$(\textit{g})    &$V_{\rm r}$ (km s$^{-1}$)& $V\sin(i)$ (km s$^{-1}$)& [M/H]& $\chi^2$\\
\hline
HD 53929& 12764 $\pm$ 200 & 12552$\pm$ 380$^a$ &3.71 $\pm$ 0.20&14.2 $\pm$ 2.1&25.8 $\pm$ 1.9&-1.0 & 2.20 \\
HD 63975& 12089 $\pm$ 200 &  12283$\pm$  380$^a$ &3.27 $\pm$ 0.20&32.8 $\pm$ 1.6&28.7 $\pm$ 0.9&-0.5 & 2.06 \\
\hline
	\end{tabular}
\end{center}
{\it Notes:}{$^a$data obtained from the [$c_{\rm 1}$]-photometric temperature calibration (Napiwotzky~et~al. \citeyear{Napiwotzky+93}) and corrected according to Netopil~et~al. (\citeyear{Netopil+08})}.
\end{table*}

For both stars, the values of $T_{\rm eff}$ obtained in this study from the fitting of Balmer line profiles are consistent with the results derived from the photometric temperature [$c_{\rm 1}$]-calibration (Napiwotzky~et~al. \citeyear{Napiwotzky+93}) taking into account the  correction proposed by Netopil~et~al. (\citeyear{Netopil+08}) for the HgMn stars
(see Table~\ref{tab:velocities}). To carry out the photometric temperature [$c_{\rm 1}$]-calibration for HD~53929 and HD~63975 we have used the results of $uvby$ photometry published respectively by Hauck~\&~Mermilliod (\citeyear{Hauck+Mermilliod98}) and Napiwotzky~et~al. (\citeyear{Napiwotzky+93}). Then the obtained values of effective temperature are corrected employing the empirical formula derived by Netopil~et~al. (\citeyear{Netopil+08}) for the CP3 and CP4 stars.
The errors given in Table~\ref{tab:velocities} are obtained by taking into account the precision of Str\"{o}mgren photometry given in Hauck~\&~Mermilliod (\citeyear{Hauck+Mermilliod98}) and in Napiwotzky~et~al. (\citeyear{Napiwotzky+93}), and the errors for each component in the correction formula (Netopil~et~al. \citeyear{Netopil+08}).

In the case of HD~53929, our estimation of surface gravity is close to the value of $\log$(\textit{g}) = 3.60$\pm$0.25 derived by Smith~\&~Dworetsky (\citeyear{Smith+Dworetsky93}) for this star (see Table~\ref{tab:velocities}). Meanwhile, the obtained value of $T_{\rm eff}$ is significantly smaller than the ones reported by Khalack~\&~LeBlanc (\citeyear{Khalack+LeBlanc15b}) and by Smith~\&~Dworetsky (\citeyear{Smith+Dworetsky93}). This discrepancy can be explained by the fact that Smith~\&~Dworetsky (\citeyear{Smith+Dworetsky93}) have adopted $T_{\rm eff}$ and $\log$(\textit{g}) as the mean values of estimates derived from the analysis of Str\"{o}mgren photometry data (Hauck~\&~Mermilliod \citeyear{Hauck+Mermilliod80}) using the ($c_0$, $\beta$) grids of Moon~\&~Dworetsky (\citeyear{Moon+Dworetsky85}) and from the "best fit" of only one Balmer line profile $H_{\gamma}$. Meanwhile, to fit the Balmer line profiles in the spectrum of HD~53929, Khalack~\&~LeBlanc (\citeyear{Khalack+LeBlanc15b}) have mistakenly used a grid of normalised theoretical fluxes, for which the FITSB2 code (Napiwotzki \citeyear{Napiwotzki+04}) usually results in overestimated values of $T_{\rm eff}$ and $\log$(\textit{g}). The effective temperatures and gravities published by Khalack~\&~LeBlanc (\citeyear{Khalack+LeBlanc15b}) for the other stars in their sample were obtained using the grids of non-normalised theoretical fluxes.

\begin{table}
\begin{center}
	\caption{Average abundances obtained for HD 53929 and HD 63975.}
\def\arraystretch{0.96}
\setlength{\tabcolsep}{3pt}
	\label{tab:abundances}
	\begin{tabular}{l|rcc|rcc}
		\hline
		El.& \multicolumn{3}{c|}{HD~53929 : [X/H] } & \multicolumn{3}{c}{HD~63975 : [X/H] }   \\
		   &     $N$       &    Here  &  Other &    $N$   &  Here  & Other \\
           &               &               & studies &        &               & studies \\
		\hline
		C\,{\sc ii} &  0  &  -              & - & 1  &  0.15             & -      \\
		O\,{\sc i}  &  4  &  -1.33$\pm$0.64 & - & 8  &  -0.55 $\pm$ 0.42 & -   \\
		Ne\,{\sc i} &  0  &  -              & - & 4  &  2.97 $\pm$ 0.29  & -  \\
		Mg\,{\sc ii}&  2  &  -0.97$\pm$0.23 &  -0.94$\pm$0.18$^a$ & 6  &  -0.65$ \pm$ 0.08 & -      \\
		Si\,{\sc ii}&  9  &   0.13$\pm$0.12 &  0.04$\pm$0.07$^a$ & 6  &  -0.42 $\pm$ 0.40 & -   \\
		P\,{\sc ii} & 20  &   2.01$\pm$0.23 & - &13  &  1.86 $\pm$ 0.31  & -  \\
		S\,{\sc ii} &  4  &  -0.18$\pm$0.50 & - & 0  &  -                & -  \\
		Ti\,{\sc i} &  0  &  -              & - & 4  &  -0.33 $\pm$ 0.71 & -      \\
		Ti\,{\sc ii}&  3  &  0.77$\pm$0.31  & - & 5  &  0.02 $\pm$ 0.13  & -    \\
		Cr\,{\sc i} &  0  &  -              & - & 2  &  0.67 $\pm$ 0.52  & -    \\
		Cr\,{\sc ii}&  2  &  2.13$\pm$0.86  & -1.22$\pm$ 0.10$^b$ & 0  &  -  & -  \\
		Mn\,{\sc i} &  2  &  1.01$\pm$0.33  & - & 2  &  1.15 $\pm$ 0.31  & -    \\
		Mn\,{\sc ii}&  7  &  1.34$\pm$0.86  &  0.73$\pm$0.20$^b$ & 7  & 1.31 $\pm$ 0.63  & -     \\
		Fe\,{\sc ii}& 28  &  0.73$\pm$0.31  &  0.23$\pm$ 0.05$^b$ &80  & 0.57 $\pm$ 0.24  & -    \\
		Hg\,{\sc ii}&  0  &  $\leq$ 2.5     &  0.93$\pm$0.10$^c$ & 1  &  4.66  &   2.94 $\pm$ 0.14$^d$  \\
		\hline
	\end{tabular}
\end{center}
{\it Notes:}{ $^a$Smith (\citeyear{Smith1993}), $^b$Smith \& Dworetsky (\citeyear{Smith+Dworetsky93}), $^c$Smith (\citeyear{Smith1997}), $^d$ Woolf \& Lambert (\citeyear{Woolf+Lambert99}).}
\end{table}

We then analyzed the line profiles of detected ions using the modified ZEEMAN2 code (Khalack \& Wade \citeyear{Khalack+Wade06}) developed by Landstreet (\citeyear{Landstreet88}). Knowing that HgMn stars are generally non-magnetic stars, all parameters related to the magnetic field structure are set to zero in our model. By fitting the synthetic profile with the observed one, the code determines three free model parameters: the chemical abundance of the ion, the radial velocity and the rotational velocity of the star.

For each line, the simulation is repeated several times in order to reach the absolute minimum in the space of free model parameters and to get closer to real values of chemical abundance of the studied ion, radial velocity $V_{\rm r}$ and rotational velocity $V$sin$i$ of the star. This was done by changing the initial values of the aforementioned parameters at each simulation and by using the downhill simplex method (Press et al. \citeyear{Press1992}) to find the best fit (Khalack \& Wade \citeyear{Khalack+Wade06}). A preliminary analysis of these three parameters allows us to decide which lines are well fitted and which ones may be misidentified or are blends due to a significant contribution from other ions.

The ZEEMAN2 code allows to estimate the optical depth $\mathit{\tau}_{l}$ at which the core of a spectral line is formed (Khalack et al. \citeyear{Khalack+07}, \citeyear{Khalack+17}).
For each analysed line profile the code determines the abundance of studied element and the optical depth $\mathit{\tau}_{l}$ of the line core formation. Then it finds the respective optical depth $\mathit{\tau}_{5000}$ of standard scale (calculated for $\lambda=5000$ \text{\AA}).
In this method, the standard scale of optical depth $\mathit{\tau}_{5000}$ is specified according to the PHOENIX model (Hauschildt et al. \citeyear{Hauschildt+97}) for all layers of the stellar atmosphere.

Thus, when a fairly large number of line profiles is detected for a given ion, and assuming that their cores are formed at different depths, it is then possible to probe the atmosphere for any variation in abundance as a function of optical depth. In other words, one can verify if element's abundance is vertically stratified in the stellar atmosphere of the studied star. The slope of the abundance change is considered to be statistically significant if its measured value exceeds its uncertainty more than three times.
When this condition is satisfied and the abundance of an element varies strongly (by 0.5 dex or more) in the probed layers, we may then conclude that a vertical stratification is detected for that element. This method of analysis was first applied to detect the vertical stratification of certain chemical elements including iron (Khalack et al. \citeyear{Khalack+07}, \citeyear{Khalack+08}, \citeyear{Khalack+10}) in stellar atmospheres of blue horizontal-branch stars (i.e. helium-burning stars). Theoretical models such as Hui-Bon-Hoa et al. (\citeyear{HBH+00}) and LeBlanc et al. (\citeyear{LeBlanc+09}) also predict the stratification of some elements in these stars. In our simulations the estimation errors are caused mainly by the uncertainties present in the atomic data, in the observation data and in the calculation of the theoretical models. Taking into account that these uncertainties are not easy to measure, the requirement for the strong variation of abundance (0.5 dex) is imposed to ensure that the detected abundance trend does not represent an arbitrary variation due to these uncertainties.

\subsection{Radial and rotational velocities}
\label{radial}
The averages of the radial and rotational velocities derived from the spectral lines selected for each star are shown in Table~\ref{tab:velocities}. For HD~53929, the values of the radial velocity 13.3 km s$^{-1}$ obtained by Zentelis (\citeyear{Zentelis+83}) and 15 $\pm$ 1 km s$^{-1}$ derived by Khalack \& LeBlanc (\citeyear{Khalack+LeBlanc15b}) are in good agreement with our result. The value 6.1 km s$^{-1}$ derived by Evans (\citeyear{Evans67}) and Hube (\citeyear{Hube70}), and the value 11.2 $\pm$ 1.3 km s$^{-1}$ obtained by Gontcharov (\citeyear{Gontcharov+06}) for HD 53929 are lower than ours and may suggest that this star may be a member of a long periodic binary system. The average rotational velocity found in this study for HD~53929 is also consistent with the values of 30 km~s$^{-1}$ and 25 km s$^{-1}$ previously found respectively by Smith \& Dworetsky (\citeyear{Smith+Dworetsky93}) and by Royer et al. (\citeyear{Royer+2002}) taking into account the simulation errors. For HD~63975, the radial velocity value of 32.3~km s$^{-1}$ obtained by Wilson (\citeyear{Wilson+53}) is in good agreement with the data presented in Table~\ref{tab:velocities}. The value of 30 km s$^{-1}$ found by Wolff \& Preston (\citeyear{Wolff+Preston78}) for the rotational velocity of HD~63975 is also consistent with our result.

\begin{table}
	\caption{The list of spectral lines used for the abundance analysis of O\,{\sc i} in the spectrum of HD 53929. The complete list of spectral lines used for the abundance analysis of both stars can be found online.}
	\label{tab:lines}
	\begin{tabular}{ l l l c c}
		\hline
		Ion&$\lambda$(\AA)&log $N_{\rm X}/N_{\rm tot}$ &log ($gf$)&$E_{\rm i}$ (cm$^{-1}$)
		\\\hline
		O\,{\sc i}&6158.187&-5.392 $\pm$ 0.001&-0.409&86631.45
	  	\\
      	O\,{\sc i}&7771.940&-3.916 $\pm$ 0.007& 0.369&73768.20
      	\\
      	O\,{\sc i}&7775.388&-4.445 $\pm$ 0.013& 0.002&73769.08
      	\\
      	O\,{\sc i}&8446.360&-4.171 $\pm$ 0.012& 0.236&76794.98
		\\\hline
  \end{tabular}
\end{table}

\begin{figure*}
	\begin{tabular}{cc}
		\includegraphics[scale=0.3,angle=-90]{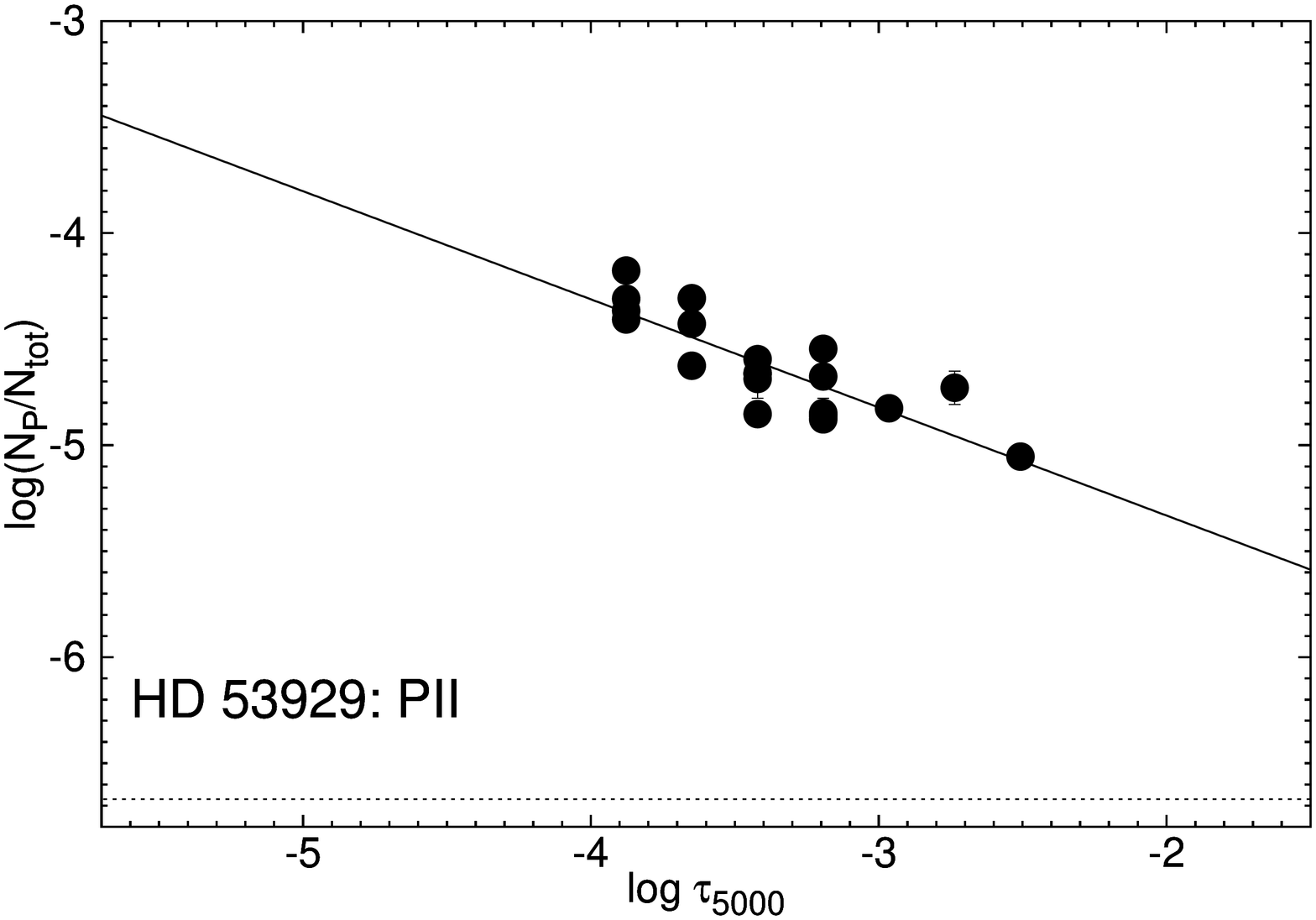} &
        \includegraphics[scale=0.3,angle=-90]{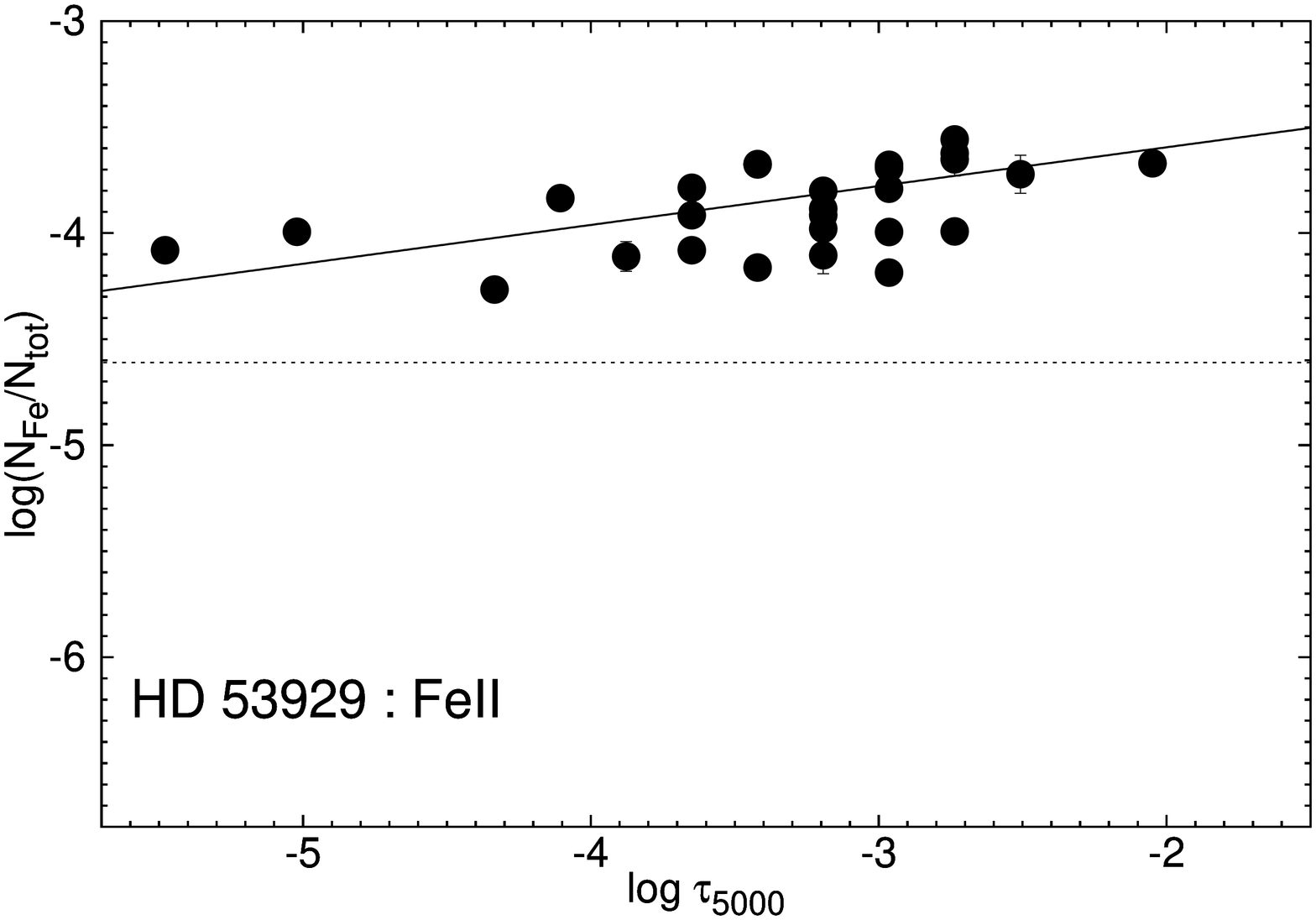}  \\
	\end{tabular}
	\caption{The abundance of P and Fe relative to the total number of atoms obtained from individual lines for HD~53929 under consideration as a function of optical depth at 5000 \text{\AA}. The circles represent the singly ionized ions and the triangles the neutral ions. The dashed line represents the solar abundance of the chemical element with respect to the $N_{tot}$. 
Linear fits of the abundance stratification are also shown (solid lines).}
	\label{fig:strat1}
\end{figure*}

\begin{figure*}
	\begin{tabular}{cc}
		\includegraphics[scale=0.3,angle=-90]{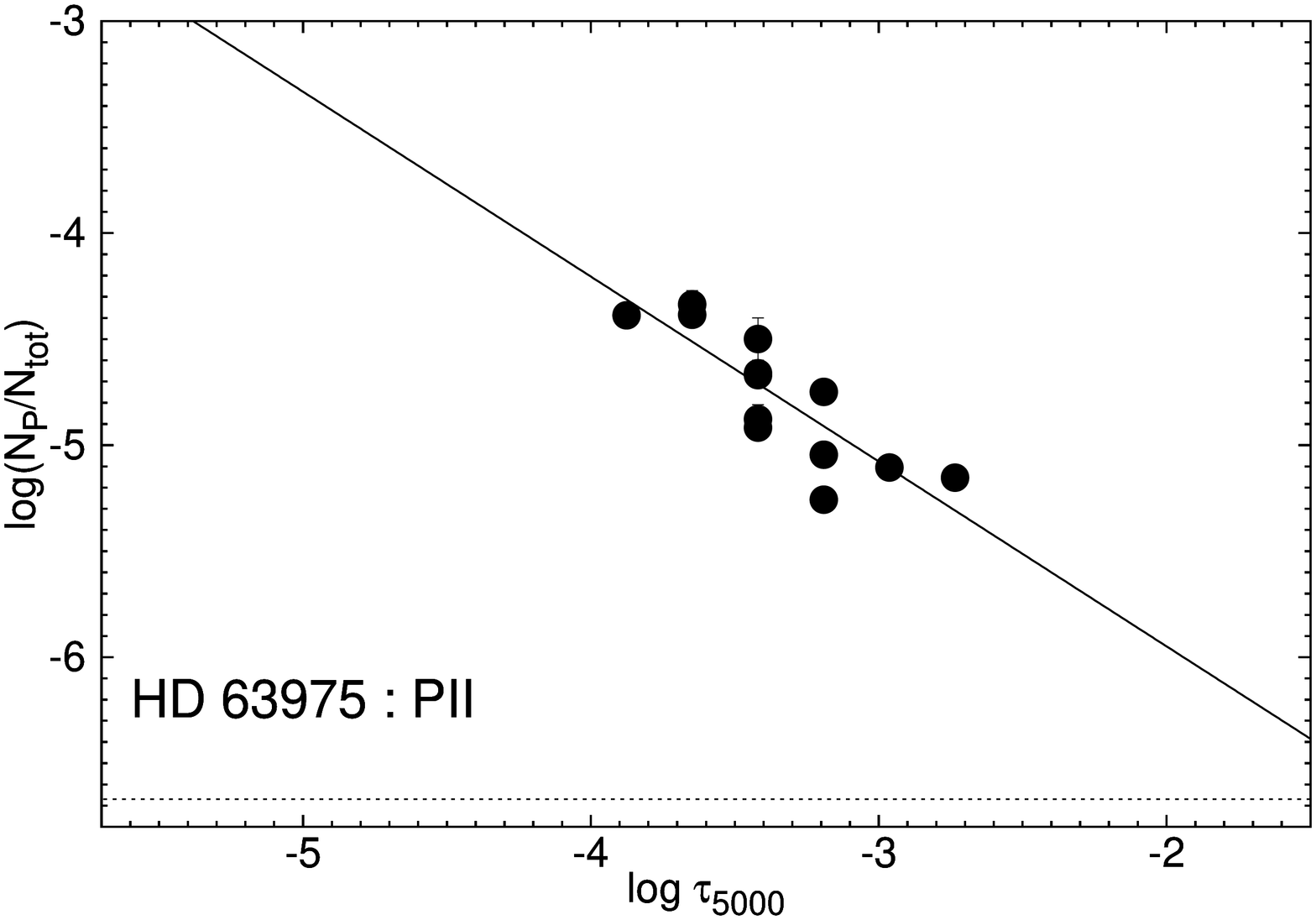} &
        \includegraphics[scale=0.3,angle=-90]{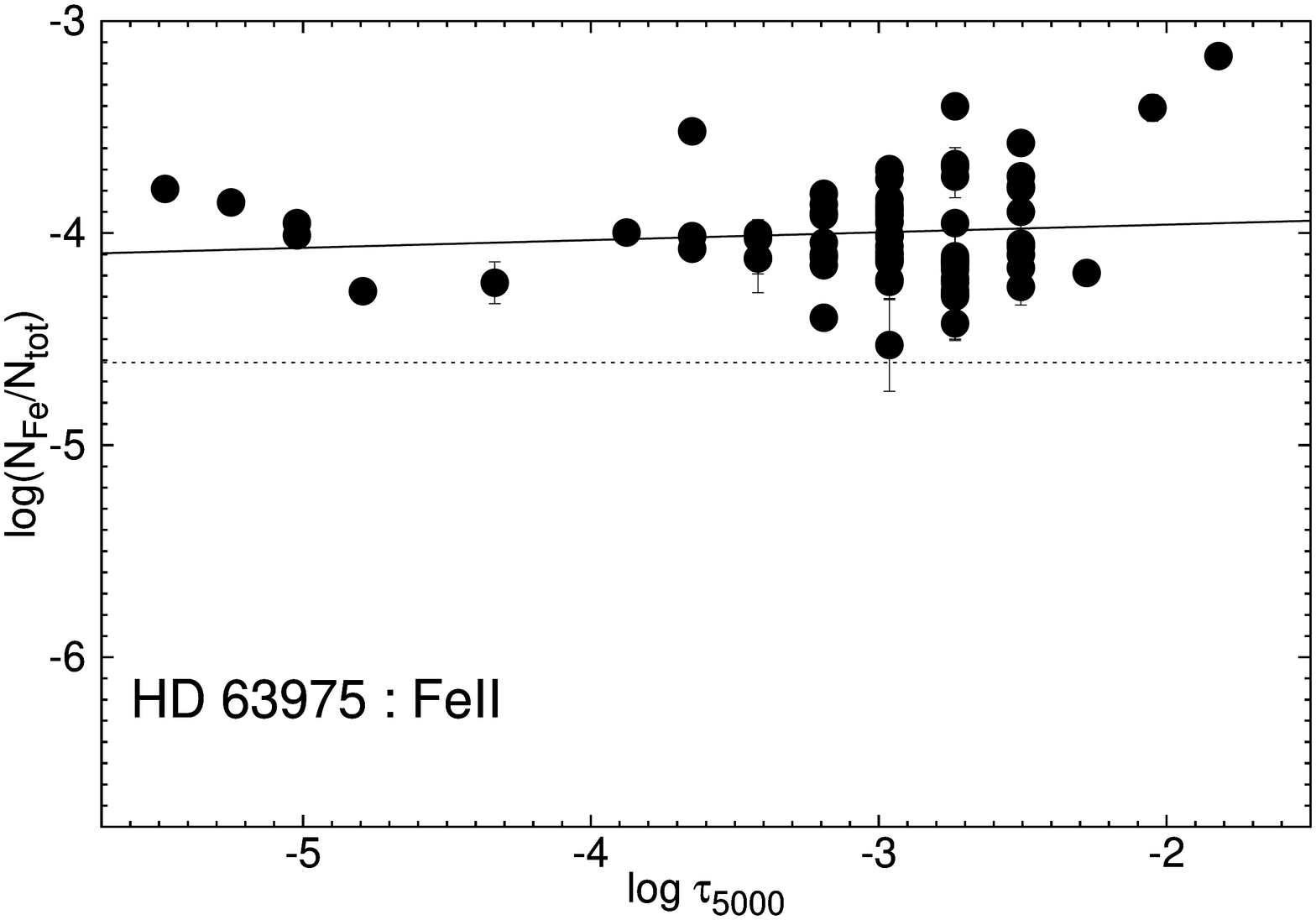} \\
	\end{tabular}
	\caption{The abundance of P and Fe relative to the total number of atoms obtained from individual lines for HD~63975.}
	\label{fig:strat2}
\end{figure*}

\subsection{Abundance analysis}
\label{analysis}
Table~\ref{tab:abundances} presents results derived from the analysis of absorption lines profiles found in the obtained spectra of HD~53929 and HD~63975. For each ion, we show the derived average abundance relative to its solar value (Grevesse et~al. \citeyear{Grevesse+10}, \citeyear{Grevesse+15}; Scott et~al. \citeyear{Scott+15a}, \citeyear{Scott+15b}), where $N$ represents the number of selected lines of the studied ion. 
The estimation errors of average abundances are determined by calculating the mean squared error of the data obtained for all lines of the considered ion. Results of previous studies are also shown in Table~\ref{tab:abundances} for each reported element which were recalculated with respect to recent measurements of solar abundances (Grevesse et~al. \citeyear{Grevesse+10}, \citeyear{Grevesse+15}; Scott et~al. \citeyear{Scott+15a}, \citeyear{Scott+15b}).

In contrast to studies using the IUE spectra (Smith \& Dworetsky \citeyear{Smith+Dworetsky93}; Smith 1993, 1997), we have detected He\,{\sc i}, O\,{\sc i}, P\,{\sc i}, S\,{\sc ii}, and Ti\,{\sc ii} lines in the visible spectrum of HD~53929, but no Al\,{\sc ii}, Co\,{\sc ii}, Ni\,{\sc ii}, Hg\,{\sc i} or Hg\,{\sc ii} lines were found in our ESPaDOnS spectra. For the average abundances in this star, we found that O and Mg are underabundant, that Si and S have a nearly solar abundance, while Ti and Fe are slightly overabundant and P, Cr and Mn are largely overabundant. Even though no Hg lines are visible in our spectra, we used its 3984 \text{\AA} line to estimate its upper limit abundance at +2.5 dex relative to solar. Our results for the Mg and Si abundances are consistent to the ones found by Smith (1993). 
The Mn overabundance found here is consistent to that found by Smith \& Dworetsky (\citeyear{Smith+Dworetsky93}), while our results for Cr and Fe are different from theirs, especially Cr which differs greatly. The higher S/N ratio obtained at CFHT as compared to the IUE spectra used by Smith \& Dworetsky (\citeyear{Smith+Dworetsky93}) could partly explain these discrepancies. The use of different radiative transfer codes and model atmospheres could also be contibuting factors. On other hand, the detected discrepancies could be an indication of possible binary nature of HD~53929 as it was predicted by Zentelis (\citeyear{Zentelis+83}).
In a close binary system a weaker secondary component can contribute to the observed line profiles (see, for example, Ryabchikova~et~al. \citeyear{Ryabchikova+98}) and this contribution will vary with the phase of orbital motion.

For HD~63975, our analysis shows that O and Mg are underabundant, that C, Si and Ti are near their solar abundance (i.e. within 0.5 dex), while Cr and Fe are slightly overabundant and Ne, P, Mn and Hg show large overabundances. The Hg abundance found here is larger than the one found by Woolf \& Lambert (\citeyear{Woolf+Lambert99}). Since our analysis of Hg is solely based on a single line (3984 \text{\AA}) it is not possible to estimate an uncertainty for our value, but it could be relatively large.

Generally speaking, the average abundances derived here for two stars are consistent to those found in HgMn stars (e.g. Ghazaryan \& Alecian \citeyear{Ghazaryan+Alecian+16}). In particular, the underabundance of Mg and the overabundance of P and Mn found in these two objects confirm their classification as HgMn stars as it was suggested previously by Renson \& Manfroid (\citeyear{Renson+Manfroid09}).

However, a possible source of uncertainty in our results is related to possible NLTE effects. For example, Takeda \& Honda (2016) studied NLTE effects on the oxygen abundance of B-type stars using the O\,{\sc i} triplet at 7771 \AA. They showed that this triplet is stronger in NLTE, therefore leading to a smaller abundance from ranging from 0.6 to 1.7 dex in B-type stars. The NLTE analysis of the O\,{\sc i} triplet at 7771 \AA\, would then result in a higher deficit of oxygen in the atmospheres of the stars studied here (see Tab~\ref{tab:lines}). This example shows that NLTE effects can be important for certain atomic lines.

\subsection{Vertical abundance stratification}
\label{strat}
For some chemical species we have identified a sufficient number of line profiles that can be used to analyse a possible variation of element's abundance with optical depth in the stellar atmospheres of studied stars.

As discussed in Section~\ref{intro}, some elements such as Mn and Cr have been observed to be stratified in the stellar atmosphere of several HgMn stars. The list of spectral lines used to derive the average abundance of O\,{\sc i} in HD~53929 is presented in Table~\ref{tab:lines} as an example. The first and second columns indicate respectively the considered ion and the central wavelength of the spectral line. In the 3rd, 4th and 5th columns, we present respectively the abundance found, the oscillator strength and the associated lower energy level for each spectral line. Complete selected line lists for all ions studied are given in appendix.

The distributions of the chemical abundance of the analyzed ions as a function of the optical depth are shown in Fig.~\ref{fig:strat1} for HD~53929 and in Fig.~\ref{fig:strat2} for HD~63975, and the linear regressions obtained with the help of the software GNUPLOT (version 5.0 of January 2015, Williams \& Kelley \citeyear{Williams+86}) taking into account only the derived abundance estimates. In each graph, each point represents an abundance estimate obtained from the analysis of a single line profile. We present these graphs only for chemical species for which we have obtained abundance estimates from the analysis of at least 10 line profiles. The procedure of the determination of the optical depth of line core formation is described in details in Khalack et al. (\citeyear{Khalack+07}, \citeyear{Khalack+17}). 
The dashed line shows the level of the solar abundance of the studied chemical element. The straight line shown on each graph represents the best linear regression derived using the least squares method. For each case, the slope and its uncertainty have been calculated using the GNUPLOT software by averaging the quadratic errors between the linear regression and the data points.

Fig.~\ref{fig:strat1} shows that phosphorus appears to have a strong vertical stratification in the outer atmospheric layers of HD~53929 with a statistically significant slope (-0.51 $\pm$ 0.08) and a strong abundance variation (more than 0.5 dex). As mentioned previously, since several sources of uncertainties exist in abundance analysis (e.g. atomic data, underlying model atmospheres, NLTE effects, etc.), a strong abundance variation (of at least 0.5 dex or more) in the atmospheric layers where the selected lines are formed is required to confidently confirm the presence of stratification. Meanwhile, the slope (0.18 $\pm$ 0.07) found for Fe is not statistically significant and its abundance does not vary strongly in HD~53929 (see Fig.~\ref{fig:strat1}). Therefore, we cannot conclude that Fe is stratified in the stellar atmosphere of HD~53929.

Phosphorus also shows a clear vertical stratification in the stellar atmosphere of HD~63975 (see Fig.~\ref{fig:strat2}), where the slope is found to be statistically significant (-0.87 $\pm$ 0.17) and a strong abundance variation exists. As in the other star, Fe shows an abundance slope that is not statistically significant (0.04 $\pm$ 0.04) in the atmosphere of HD~63975.

\section{Discussion and conclusions}
\label{discus}
The aim of this paper was to perform a spectral analysis of two HgMn-type stars (HD~53929 and HD~63975) selected from the sample of Project VeSElkA, in order to verify for the presence of vertical stratification of element abundances in their atmosphere. We first identified the absorption lines present in each spectrum and then carried out simulations using ZEEMAN2 code (Landstreet \citeyear{Landstreet88}) modified by Khalack \& Wade (\citeyear{Khalack+Wade06}), which allowed us to estimate the average abundance of each detected ion along with the radial and the rotational velocity of each star. Our simulations were carried our assuming LTE, therefore NLTE effects remain a source of uncertainty.

We have analyzed absorption lines of several ions to determine their average abundance in the atmosphere of HD~53929 and HD~63975 (see Table~\ref{tab:abundances}). Oxygen and magnesium are found to be deficient in the atmosphere of the two stars, while phosphorus, chromium, manganese and iron are overabundant. Mercury is found to be strongly overabundant in HD~53929. The strong overabundance obtained for Mn and the abundances found for the other elements allow us to conclude that HD~53929 and HD~63975 are indeed HgMn stars as previously determined by Renson \& Manfroid (\citeyear{Renson+Manfroid09}).

For both stars, the values of radial and rotational velocity found in our study are consistent with previous studies (see Subsection~\ref{radial}). For HD~53929, since a certain dispersion of the radial velocity exists in comparison to the previous studies, it could indicate that HD~53929 is a binary star as predicted by Zentelis (\citeyear{Zentelis+83}). More data is needed for a firm conclusion.

After studying the abundance variation as a function of optical depth for elements with a sufficient number of detected line profiles, we found that phosphorus is vertically stratified in the atmosphere of HD~53929 and HD~63975 and that it exhibits the same behavior in both stars. Its abundance increases strongly towards the superficial layers.
Similar results for the stratification of phosphorus abundance with optical depth have been reported recently by Catanzaro~et~al. (\citeyear{Catanzaro+16}) for another HgMn star HD49606, although the position of the stratification profile is deeper in that star than those studied here. The stratification detected for phosphorus indicates that atomic diffusion could play an important role in the atmosphere of these two stars.

\section*{Acknowledgments}
Authors are thankful to the Facult\'{e} des \'{E}tudes Sup\'{e}rieures et de la Recherche de l'Universit\'{e} Moncton for the financial support of this research. F.L. and
V.K. acknowledge support from the Natural Sciences and Engineering Research Council of Canada.
Part of the calculations have been done on the supercomputer {\it briarree} of the University of Montreal, under the guidance of Calcul Qu\'{e}bec and Calcul Canada. The use of this supercomputer is funded by the Canadian Foundation for Innovation (CFI), NanoQu\'{e}bec, RMGA and Research Fund of Qu\'{e}bec - Nature and Technology (FRQNT). This work has made use of the VALD database operated at Uppsala University, the Institute of Astronomy RAS in Moscow and the University of Vienna, as well as the NIST database.	
This paper has been typeset from a \TeX/\LaTeX\, file prepared by the authors.

\label{lastpage}
\end{document}